# Using Polymer Electrolyte Gates to Set-and-Freeze Threshold Voltage and Local Potential in Nanowire-based Devices and Thermoelectrics

*Sofia Fahlvik Svensson, Adam M. Burke, Damon J. Carrad, Martin Leijnse, Heiner Linke, and Adam P. Micolich\**

The strongly temperature-dependent ionic mobility in polymer electrolytes is used to "freeze in" specific ionic charge environments around a nanowire using a local wrap-gate geometry. This makes it possible to set both the threshold voltage for a conventional doped substrate gate and the local disorder potential at temperatures below 220 K. These are characterized in detail by combining conductance and thermovoltage measurements with modeling. The results demonstrate that local polymer electrolyte gates are compatible with nanowire thermoelectrics, where they offer the advantage of a very low thermal conductivity, and hold great potential towards setting the optimal operating point for solid-state cooling applications.

## 1. Introduction

Semiconductor nanowires are a highly promising nanomaterial with diverse applications ranging from high efficiency thermoelectrics [1–3] to nanoscale transistors. [4,5] An on-going quest towards these ends is the development of improved strategies for conductivity and Fermi level control via electrostatic gating. Gating via a doped substrate [6] is simple and effective but provides no scope for local control. Nanoscale metal gates beneath [7] and above [8] a nanowire enable local control but push carriers against the nanowire surface [9] subjecting them to strong surface scattering. [10] The desire for stronger gate coupling and improved homogeneity drove the development of cylindrical wrap-gates for laterally oriented nanowires. This was first achieved using conventional metal/oxide formulations, [11,12] and more recently, using polymer electrolyte gating. [13,14] A

S. F. Svensson, Dr. M. Leijnse, Prof. H. Linke
Solid State Physics and
Nanometer Structure Consortium
(nmC@LU), Lund University
S-221 00, Lund, Sweden
Dr. A. M. Burke, D. J. Carrad, Prof. A. P. Micolich
School of Physics
University of New South Wales
Sydney NSW 2052, Australia
E-mail: adam.micolich@nanoelectronics.physics.unsw.edu.au

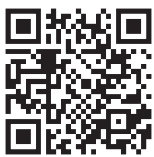

polymer electrolyte gated nanowire transistor consists of a salt-laden polymer gel, e.g., $LiClO_4$ in poly(ethylene oxide), spanning a gap between a metal gate electrode and the nanowire. A voltage applied to the gate electrode drives migration of $Li^+$ and $ClO^-_4$ ions to form electric double layers at the electrode/electrolyte and electrolyte/nanowire interfaces. [15] This effective transfer of gate charge to within ≈1 nm of the nanowire gives considerably improved gate coupling and sub-threshold characteristics. [13,14] Whereas earlier work focused on organic semiconductor transistors, [15,16] and semiconductor nanowire transistors with unpatterned polymer electrolyte films, [13] this paper presents the first study of the low temperature electrical properties of a nanowire transistor featuring a nanoscale polymer electrolyte patterned by electron beam lithography. [14]

In this paper we focus our attention on the properties and potential uses of a unique aspect of polymer electrolyte gates, namely the fact that the ionic mobility drops rapidly to zero as the temperature $T$ is reduced below approximately 220 K. [16] This enables the electrolyte's ion distribution to be set and "frozen in" to give a fixed external charge environment near the nanowire's surface. It holds interesting potential uses in quantum transport studies, which are typically performed at $T < 4$ K; for example, the ability to freeze in the ion distribution in polymer electrolyte gated nanowires has been recently used to study spin-orbit effects in nanowires. [13] Here we show that this approach can be extended to add two new features: a) the ability to tune the threshold voltage for conventional electrostatic gates, for example, an insulated, doped substrate [6] over a wide range; and b) the ability to set the disorder potential for nanowire transistors and quantum devices. The disorder potential breaks the nanowire into a string of quantum dots coupled in series, [17] with properties that can be tuned using the voltage applied to the polymer electrolyte gate. We characterize our devices using conductance and thermovoltage measurements with a key result being the demonstration that local polymer electrolyte gates are fully compatible with thermoelectric measurements of nanowires [18–25] offering the advantage of strongly reduced thermal conductivity compared to metal gates. This is of high potential interest for solid-state cooling applications, where the ability to control disorder and set an optimal operation point using a gate, may enable optimization of the expected[26,27] and recently demonstrated[24] enhancement of thermoelectric power factor in disordered nanowires compared to bulk materials.

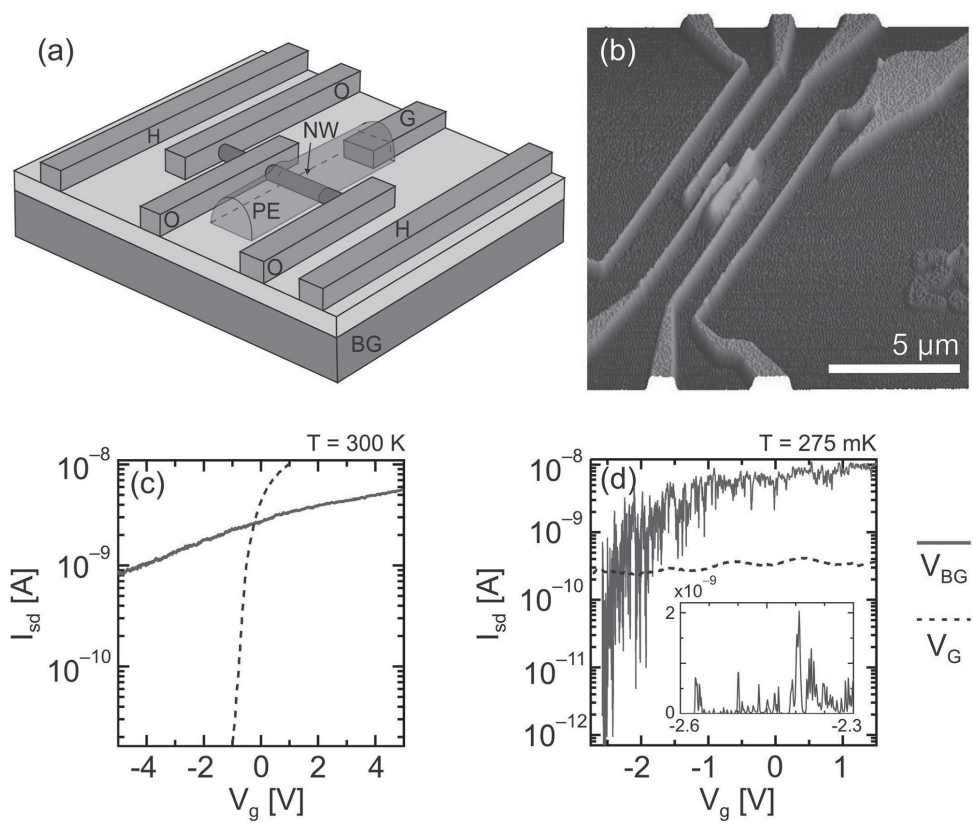

**Figure 1.** a) Schematic and b) atomic force micrograph of the device used in this study, which consists of an InAs nanowire (NW), three NiAu Ohmic contacts (O), an insulated n⁺-Si back-gate (BG), a polymer electrolyte dielectric (PE) connected to a gate electrode (G) and two heater strips (H) for applying a thermal gradient along the nanowire. c,d) Source-drain current $I_{sd}$ vs gate voltage $V_g$ at temperatures c) $T = 300$ K and d) $T = 0.275$ K. In each, the solid trace is the back-gate voltage $V_g = V_{BG}$ and the dashed trace is the PE gate voltage $V_g = V_G$, with the other gate held at ground. In (c) we have $V_{PE} = V_g$ whereas in (d) $V_{PE}$ is fixed at 0 V due to ionic freeze-out. The inset to (d) is a focus on $I_{sd}$ at most negative $V_{BG}$.

## 2. Results and Discussion

**Figure 1**a,b shows a schematic and atomic force micrograph of our device. It consists of a 50 nm diameter InAs nanowire, three NiAu Ohmic contacts (O), two NiAu heater strips (H), and a patterned PEO/LiClO₄ polymer electrolyte (PE) connected to a NiAu gate electrode (G). The behavior we report arises from the polymer electrolyte, and we would expect it to hold for semiconducting nanowires of other compositions and diameters. The device is made on an insulated, degenerately doped Si substrate, which acts as a back-gate (BG). The Ohmic contacts, heater strips and gate electrode are deposited in a single electron-beam lithography and metal deposition step, with the electron-beam patterned polymer electrolyte processed thereafter.[14] The electrical studies consist of two sets of measurements via contacts O1 and O2: the ac current $I_{sd}$ passing through the nanowire in response to an applied source-drain voltage $V_{sd}$ and the open-circuit dc thermovoltage $V_{th}$ in response to an applied heater voltage $V_H$ on H1, resulting in a thermal gradient $\Delta T$ along the nanowire. The source-drain voltage consists of a base ac excitation of 100 μV at 13 Hz to facilitate low-noise phase-sensitive detection and an optional dc component <±5 mV for producing Coulomb blockade stability diagrams.[28]

The PE gate's ionic nature gives it a "set and hold" capability made possible by the freezing in of the ion distribution in the polymer electrolyte.[16] To demonstrate this, we distinguish the voltage $V_G$ externally applied to electrode G from the potential $V_{PE}$ experienced by the nanowire due to the polymer electrolyte: $V_{PE}$ is maintained below 220 K even if the voltage $V_G$ applied to electrode G is altered. We therefore used the following measurement methodology: a) the PE gate voltage $V_{PE}$ is set at $T = 300$ K with back-gate voltage $V_{BG} = 0$; b) the device is cooled to $T = 0.275$ K; c) measurements of $I_{sd}$ and $V_{th}$ versus $V_{BG}$ are obtained; d) the device is warmed to $T = 300$ K where $V_{PE}$ can be adjusted for the next series of measurements. This four-step process was repeated for several $V_{PE}$ settings between −1 V and +1 V. At the end of Step b) we run $V_G$ down to zero and keep G grounded to fully demonstrate the "set and hold" capability of the PE gate; equivalent results are obtained if G is held at $V_G = V_{PE}$. The voltage $V_G$ is returned to $V_{PE}$ immediately prior to Step d). The use of $V_{BG}$ in this device is also enabled by the frozen ions which cannot redistribute to screen changes in substrate potential at low $T$.

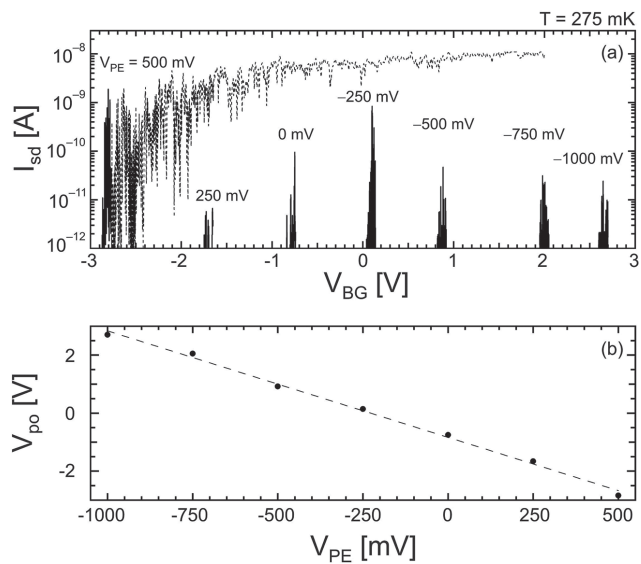

**Figure 2.** a) $I_{sd}$ vs $V_{BG}$ for the first 100 mV of $V_{BG} > V_{po}$ for the seven different $V_{PE}$ settings indicated. Additionally, the full $V_{BG}$ range for the gate trace at $V_{PE} = 500$ mV is shown as the faded blue dashed line in (a). To set $V_{PE}$, the device is warmed to room temperature, the $V_{PE}$ is changed, and the device is cooled back to $T = 0.275$ K. We define the pinch-off voltage $V_{po}$ as the first $V_{BG}$ where $I_{sd} > 10$ pA if starting from highly negative $V_{BG}$. b) Plot of $V_{po}$ vs $V_{PE}$ demonstrating the ability to use the PE gate to set the threshold for the underlying n+-Si substrate back-gate.

## 2.1. Comparison Between Back-gate and Polymer Electrolyte Gate Performance

We begin by comparing the performance of the PE gate and back-gate at room and cryogenic temperatures. Figure 1c shows $I_{sd}$ versus $V_G$ (dashed) and $V_{BG}$ (solid) at $T = 300$ K; each is swept with the other gate grounded, hence the crossing of the traces at $V_G = V_{BG} = 0$. The PE gate coupling is considerably stronger, as noted in prior work.[13,14] This situation is reversed at $T = 0.275$ K (Figure 1d) where the ions in the PE gate are frozen. The back-gate drives the nanowire to pinch-off ($I_{sd} = 0$) over a much reduced bias range whereas changes in the voltage $V_G$ applied to G have little effect; the frozen ionic mobility means that $V_{PE}$ remains fixed at $T = 0.275$ K despite any change in $V_G$. The abundant fluctuations in $I_{sd}$ in the solid trace in Figure 1d arise due to quantum confinement and, in the low $I_{sd}$ limit, Coulomb blockade, as highlighted by the inset to Figure 1d. Some weak quantum interference oscillations are also apparent in the dashed trace in Figure 1d. This reflects some direct electric field coupling between the nanowire and gate electrode[14,29] rather than a change in ion distribution, that is, $V_{PE}$.

## 2.2. Demonstration of the Low-temperature "Set-and-Hold" Capability for the PE Gate

We now demonstrate how the PE gate can be used in set-and-hold mode to set the threshold voltage for the back-gate. We do this by sweeping from strongly negative $V_{BG}$ towards more positive $V_{BG}$, for each of seven different $V_{PE}$ settings (**Figure 2a**), looking for the conduction onset. We quantify onset using the pinch-off voltage $V_{po}$, which we define as the first $V_{BG}$ where $I_{sd}$ exceeds 10 pA if sweeping from maximally negative $V_{BG}$ towards positive $V_{BG}$. We show only the first 100 mV of back-gate characteristic for each case in Figure 2a to minimize clutter, as each trace individually looks qualitatively similar to the solid trace in Figure 1d overall (see the transparent dotted-line extension of the $V_{PE} = 500$ mV trace in Figure 2a). Figure 2b shows a plot of $V_{po}$ versus $V_{PE}$ for the data in Figure 2a demonstrating that $V_{po}$ can be tuned from −2.9 V to +2.6 V as $V_{PE}$ is made more negative, with a linear dependence on $V_{PE}$. The two gates counteract one another—a more positive $V_{PE}$ leads to a more negative $V_{po}$ for the back-gate. The explanation for the behavior in Figure 2 is as follows. At the most positive $V_{PE}$ there is an excess of Li+ ions at the nanowire surface. This increases the initial electron density $n$ within the nanowire at cooldown. The device is cooled with $V_{BG} = 0$ and thus a negative $V_{BG}$ is required to counteract the density enhancement due to the PE gate and shut down conduction, producing a negative $V_{po}$. The ion-induced electron density enhancement decreases for reduced $V_{PE}$ and ultimately becomes a depletion for $V_{PE} < 0$ due to an accumulation of ClO$_4^-$ ions at the nanowire surface. Here a positive $V_{BG}$ is required to reach conduction onset. By mapping how $V_{PE}$ sets $V_{po}$ (Figure 2b), one can set and 'freeze in' the operating back-gate characteristic including setting a preferred back-gate threshold voltage. The pinch-off voltage can likely be shifted well beyond the 6 V range we demonstrate here, and will ultimately be limited by insulator breakdown. The ability to tune $V_{po}$ with $V_{PE}$ is of particular interest to us because it may be highly useful for setting a nanowire to its maximal power-factor point[24] for thermoelectric and solid-state cooling applications.

## 2.3. Use of the PE Gate to Tune the Effective Disorder Potential for Transport

We now show that the PE gate can be used to enact a tuning of the effective disorder potential experienced by electrons traversing the nanowire. In **Figure 3**a–f we look at the conduction onset for six $V_{PE}$ settings between +0.5 and −1 V. In each case we plot $I_{sd}$ versus $V_{BG}$ at $T = 0.275$ K (solid black), 1.5 K (dashed mid-gray) and 2.5 K (dotted light gray). Coulomb blockade (CB) oscillations appear in all six panels but their regularity, spacing and stability is heavily dependent on $V_{PE}$. At positive $V_{PE}$ (Figure 3a–c) the CB oscillations are highly irregular, with long stretches of $I_{sd} \equiv 0$ between single peaks or small peak clusters, and are reminiscent of stochastic CB in series coupled quantum dots[30,31] and nanowires.[17] The CB structure is relatively unstable at most positive $V_{PE}$, with peaks shifting significantly between traces – this is evident by comparing traces at different $T$, which should only differ in the height and width of a given CB peak. As $V_{PE}$ is made more negative the CB oscillations become far less reproducible and matching traces can be obtained several days apart providing $V_{BG}$ is handled carefully and not swept well away from a given $V_{BG}$ range. This link between stability and $V_{PE}$ may indicate some interesting physics related to InAs surface-state trapping in strongly cationic/anionic environments, but could also be due to trap fluctuations in the back-gate oxide; further studies will be needed to establish the exact cause.

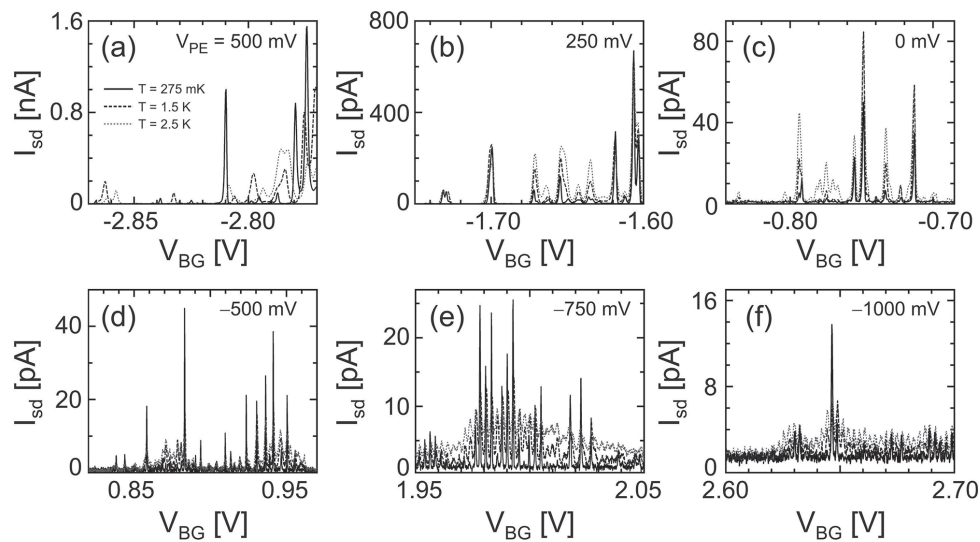

**Figure 3.** a–f) $I_{sd}$ vs $V_{BG}$ for $T$ = 275 mK (solid black), 1.5 K (dashed mid-gray), and 2.5 K (dotted light gray) for six different $V_{PE}$ settings: a) $V_{PE}$ = 500 mV, b) 250 mV, c) 0 mV, d) −500 mV, e) −750 mV, and f) −1000 mV. The traces show an evolution from highly stochastic Coulomb blockade (CB) at most positive $V_{PE}$ to more regular, periodic CB oscillations at more negative $V_{PE}$.

As $V_{PE}$ is decreased from +0.5 V to −1.0 V the CB oscillations become far more regular, particularly for $V_{PE} \leq$ −0.5 V (Figure 3d–f). Additionally, the CB peak width becomes narrower and the $I_{sd}$ at the CB maxima decreases by two orders of magnitude. The latter in particular reflects the formation of stronger, more defined tunnel barriers within the nanowire. The trend in Figure 3 indicates that changes in the ion distribution surrounding the nanowire drive a gradual transition from either a highly disordered channel featuring a long series of small, randomly sized quantum dots (QD) at positive $V_{PE}$ to a larger double or single quantum dot (QD) at negative $V_{PE}$ (see **Figure 4** and associated discussion). An interesting question is: How does the PE-gate facilitate the clear QD-like behavior that emerges with more negative $V_{PE}$ given there is no intentional source of 0D confinement during nanowire growth or device fabrication? The most obvious source of tunnel barriers are the source and drain contacts themselves; these are produced by the non-homogeneity of the back-gate electric field near the contacts, which in turn causes reduced electron density local to the contact edges.[32] But this only explains two of at least three and possibly many tunnel barriers needed to generate the CB data in Figure 3. We argue that the remainder arise from disorder in the following way. The high density of surface states in InAs nanowires leads to a high electron density near the nanowire surface.[33] Recent work has shown that repulsive impurities scatter carriers much more strongly than attractive impurities in nanowires and can act like a potential barrier with a height many times $k_B T$.[34] One possible mechanism is that at negative $V_{PE}$, the $ClO_4^-$ ion accumulation at the nanowire surface pushes carriers away from the nanowire surface. This may both reduce the screening of, and increase exposure of carriers to, negatively charged impurities in the nanowire. We hypothesize that the net result is that one or two negatively charged impurities dominate the conductance providing the tunnel barriers that enable the clear periodic CB oscillations in Figure 3d–f. Repulsive scattering from the negative ions at the nanowire surface may also contribute. In contrast, for positive $V_{PE}$, the Li+ accumulation at the surface will pull carriers closer to the surface where transport is subject to a higher density of attractive scatterers but also potentially a lower density of more effectively screened repulsive scatterers. The net result would be a greater number of weaker barriers, and characteristics that look more like a long series of small, randomly sized QDs as in Figure 3a–c.[17,30]

## 2.4. Characterization of the Device Formed using Combined Electric and Thermoelectric Measurements

To characterize the nature of the potential created by the PE gate in more detail, we now combine electric and thermoelectric measurements, which offer complementary information. We chose to work at $V_{PE}$ =−500 mV (the same as in Figure 3d), where we obtain stable, reproducible CB oscillations over timescales exceeding several days. In Figure 4a–c we show three sets of measurements obtained over the same 150 mV $V_{BG}$ range. Figure 4a presents the differential conductance $g = dI_{sd}/dV^{ac}_{sd}$, where $dV^{ac}_{sd}$ is the ac component of the applied source-drain bias $V_{sd}$, versus $V_{BG}$ at three different temperatures. Figure 4b is a CB stability diagram[35] where we plot $g$ as the color axis versus dc source-drain bias $V^{dc}_{sd}$ and $V_{BG}$; black, light gray and mid-gray correspond to low, moderate and high $g$ (shades chosen for accentuated contrast). Finally, Figure 4c presents $V_{th}$ versus $V_{BG}$ at three different $V_H$ values. The application of $V_H$ gives rise to both a global heating effect and a small $\Delta T$ that we use to generate a thermovoltage. By increasing $V_H$ we increase the device temperature and can study the thermovoltage as a function of $T$ without heating the cryostat. We characterize the relationship between $T$ and $V_H$ by looking at how the CB peak amplitude decays with $T$ at $V_H$ = 0 and with $V_H$ at $T$ = 0.275 K. We find that $V_H$ = 20, 60 and 100 mV correspond to approximate average temperatures $T$ = 0.7, 1.3 and 2.0 K, respectively.

We now examine Figure 4a,b more closely with a focus on a detailed understanding of the transport data to ensure that the thermovoltage measurements are sensible. The most notable feature of Figure 4a is that the CB peaks exhibit two different $T$ dependences. Some peaks, e.g., that marked A, become narrower and sharper as $T$ is reduced, other peaks, e.g., the series marked B, grow with increasing $T$. Such variations in the $T$ dependence have been discussed in the context of classical versus quantum Coulomb blockade.[36] In multi-level QDs with varying coupling to the leads both types of $T$ dependencies can be observed in the same QD.[37] In our case, we suggest that similar physics arise due to the formation of a serially coupled small and large QD forming in the nanowire—see inset in Figure 4d—in which case the effective lead couplings depend on whether the QD levels are aligned or not. We now turn to the Coulomb diamonds in Figure 4b for further evidence of this interpretation. Starting at small $V^{dc}_{sd}$, the most vital feature is a series of small black diamonds. Assuming transport via a serial double QD these would correspond to blockaded transport through the larger QD. The diamonds vary in size reflecting a varying level spacing and coupling to the leads. This is typical of CB in the few electron limit[36] and common in nanowire QDs.[38] The diamond vertices at $V^{dc}_{sd} = 0$ generally align with CB peaks in Figure 4a, as expected, but the diamonds do not always close (see white arrows in Figure 4b). Missing vertices indicate inhibited transport via QD levels at $V^{dc}_{sd} \cong 0$. The presence/absence of vertices is closely linked to the CB peak $T$ dependence in Figure 4a; the CB peaks corresponding to non-closing diamonds are strongly suppressed at $T = 275$ mK but grow with increasing $T$. Shifting our attention to larger $V^{dc}_{sd}$, the most visually striking feature is the mid-gray banding indicating high $g$ when $V^{dc}_{sd}$ is sufficient to overcome blockade. The onset of the mid-gray banding forms an additional pattern of larger diamonds, highlighted by the white dashed lines, corresponding to charge addition to the smaller QD. Electrons are only transported through the double QD system at small $V^{dc}_{sd}$ when the $V^{dc}_{sd} = 0$ diamond vertices of both QDs coincide. This produces the large CB peaks in Figure 4a that decrease in height with increasing $T$. Transport is otherwise suppressed at small $V^{dc}_{sd}$ because electrons cannot tunnel between the large and small QD due to the misalignment in QD levels. This produces the small CB peaks and missing vertices in Figure 4a,b. These CB peaks grow with increasing $T$ because thermal activation enables electrons to access higher QD levels. Much of the low $V_{BG}$ structure in Figure 4b bears strong resemblance to that in heterostructure-defined nanowire double QDs,[39] and one could consider the envelope of the $T = 2.5$ K trace in Figure 4a as the corresponding CB oscillations for the smaller QD. Their large broadness is consistent with a strong coupling between the small QD and its adjacent lead.

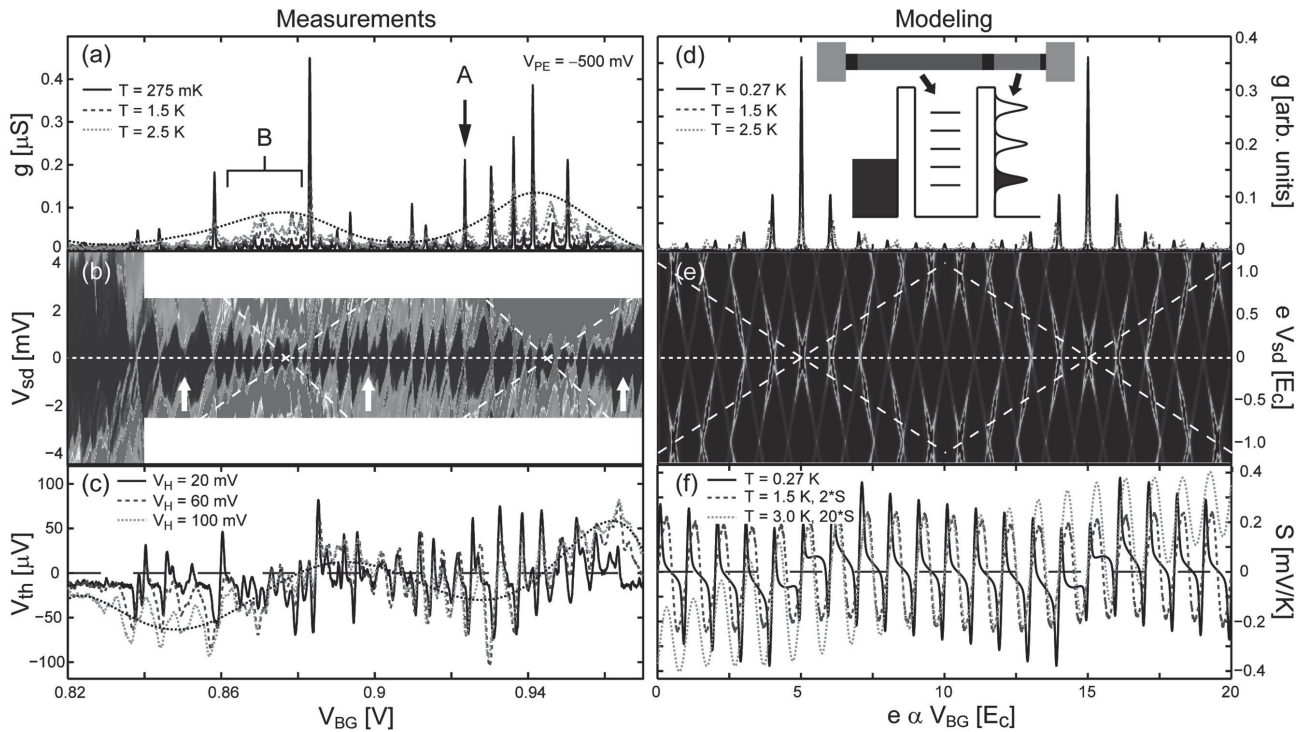

**Figure 4.** Electric/thermoelectric measurements (a–c) and associated theoretical modeling (d–f) for our NW device with $V_{PE} = -500$ mV. a) Differential conductance $g$ vs $V_{BG}$ for three different temperatures; this data matches that in Figure 3d. A and B point to regions where the peak $g$ decreases/increases with increasing $T$, similar to quantum and classical CB. The finely-dotted line is an approximate envelope for the CB peaks at $T = 2.5$ K. b) CB stability diagram showing $g$ (color axis) vs the dc source-drain bias $V^{dc}_{sd}$ ($y$-axis) and $V_{BG}$ ($x$-axis). The color axis runs from black ($g = 0$) through light gray to mid-gray ($g \cong 1$ μS) to best accentuate contrast. The data indicates that the NW splits into two adjacent dots – one large and the other small and strongly coupled to a reservoir. The small black diamonds correspond to CB in the larger dot, the larger diamonds indicated by the white dashed lines correspond to CB in the smaller dot. The white dotted line indicates $V_{sd} = 0$ and the white arrows indicate small diamonds which fail to properly close (see text). c) Open circuit dc thermovoltage $V_{th}$ vs $V_{BG}$ for three heater voltages $V_H$. The finely-dotted black line is a $7^{th}$ order polynomial fit to the $V_H = 100$ mV data to highlight the background trend (see text). d) Calculated $g$ vs effective gate voltage $e$ for the three $T$ values in (a). The inset is a schematic of the theoretical model, which consists of a nanowire quantum dot connected to a conventional reservoir on one side and a reservoir with a modulated density of states on the other. e) Calculated stability diagram corresponding to (b), the white dashed lines passing through the $g$ peaks correspond to those in (b). f) Calculated thermopower $S$ vs $V_{BG}$ for three temperatures $T$ (see text).

To support our serial double QD interpretation, in Figure 4d–f we present calculated data for a simplified double QD model. We use a simplified model because we want to cover a large gate voltage range, which corresponds to adding many electrons, making a real interacting double QD model computationally impossible. The broad mid-gray banding in Figure 4b indicates that the small QD is very strongly coupled to its adjacent lead, and we thus model this QD as a lead for the large QD with a periodically modulated density of states (DOS) (see inset to Figure 4d). We mimic the CB for the large QD by calculating $g$ using the Landauer-Büttiker formula assuming a ladder of equidistant energy levels, rather than treating the Coulomb interactions explicitly to maintain simplicity. Despite our focus on simplicity, the model bears a strong resemblance to the experimental data increasing our confidence in the double QD interpretation. On the basis of double QD formation we estimate charging energies $E_c = 1.4$ meV and 4.0 meV for the larger/smaller QD. For the nanowire's physical dimensions this suggests QD lengths of ≈600 nm for the larger QD and 30–200 nm for the smaller QD.

Finally we turn to the open circuit thermovoltage data in Figure 4c. For a weakly coupled QD, $V_{th}$ takes an approximately sawtooth-like form in the simplest case,[40,41] and a lineshape more similar to the energy derivative of the dot's conductance if second order tunneling effects, for example, co-tunneling, contribute to transport.[20,42,43] In Figure 4c, $V_{th}$ oscillates about zero for small $V_H$, and in most cases, an oscillation at given $V_{BG}$ can be directly mapped to a CB peak in the $T = 275$ mK trace in Figure 4a. The thermovoltage develops a slowly varying background as $V_H$ is increased—this is most evident at $V_H = 100$ mV and is highlighted by the black dotted line in Figure 4c. This background has zero crossings that roughly coincide with CB peaks of the small QD in Figure 4a, as expected.[41,43] The smaller QD $V_{th}$ contribution should also carry fine structure;[40,44] this may be the small-scale $V_{th}$ oscillations in Figure 4c that cannot be directly attributed to large QD CB peaks in Figure 4a. Focusing on the large QD momentarily, some of the small-scale structure in Figure 4c that corresponds directly to large QD CB peaks in Figure 4a shows a clear lineshape evolution with $V_H$. These $V_{th}$ oscillations develop from a lineshape like the energy derivative of the CB peaks at small $V_H$ towards a more sinusoidal form at large $V_H$.[20] This lineshape transition is also observed in Figure 4f where we plot the calculated thermopower $S = V_{th}/\Delta T$ in the limit $\Delta T \to 0$ for three different $T$ values. The development of an oscillating background for the high $T$ data in Figure 4f also directly reflects the experimental data. Our data, while consistent with Thierschmann et al.,[45] demonstrates that $V_{th}$ remains straightforward even in the limit where one QD is much larger than the other and the dot couplings to the leads differ significantly.

## 3. Conclusions

In conclusion, we have investigated the electric and thermoelectric properties of an InAs nanowire transistor featuring a nanoscale patterned polymer electrolyte gate.[14] We have shown that the drop in ionic mobility at ≈220 K enables us to 'freeze in' a specific surrounding ionic environment for the nanowire at low temperature, which can in turn be used to set the pinch-off voltage for a conventional, doped-substrate back-gate. The pinch-off voltage can be controllably shifted across a range spanning 6 V by changing the initial pre-cooldown polymer electrolyte gate voltage over a range of 1.5 V; this can likely be extended substantially, possibly to tens of volts, subject to the dielectric breakdown limits of the polymer and substrate oxide. Additionally, the PE gate can be used to set the disorder potential influencing transport through the nanowire. At positive PE gate voltage we see strongly stochastic Coulomb blockade oscillations indicative of the formation of several disorder-induced quantum dots in series in the nanowire.[17,30] As the PE gate voltage becomes more negative the CB oscillations become more periodic, and are consistent with two quantum dots coupled in series.

A final aspect we addressed was the potential for using PE gate structures as a route to implementing local gating of nanowires in thermoelectric measurements. The PE gate's threshold setting capability holds considerable promise for use in tuning a nanowire-based thermoelectric energy converter to its highest efficiency regime.[24] A concern with conventional metal/oxide gate electrodes, and particularly wrap-gate structures, for such a purpose is that they will 'short circuit' the applied thermal gradient because a metal's high electrical conductivity entails a high thermal conductivity via the Wiedemann-Franz law.[46] The negligible electronic conductivity and low phonon heat conductivity of polymer electrolytes should make them highly favorable for thermoelectric measurements, and we have demonstrated this by showing that thermovoltage measurements of a PE gated nanowire device bear strong correspondence to electrical transport measurements for the same device. Thinking more broadly, our study highlights that $V_{th}$ carries similar information content to a stability diagram and much more than a conductance trace alone. As a result, we see several potential practical advantages for using thermovoltage measurements over source-drain bias studies in characterizing nanowire quantum dots, these include: a) the $V_{th}$ signal remains strong even when the related CB peaks are small, b) the measurement time is shorter, and c) measurements do not require a large dc source-drain voltage which, for nanoscale devices, can result in large electric fields that significantly modify the electronic structure.[29] Overall, our work here demonstrates the strong potential for further applications of polymer electrolyte gates in both electric and thermoelectric transport studies of nanowire-based devices.

## 4. Experimental Discussion

*Device Fabrication*: Nanowire devices were fabricated from 3 µm long, 50 nm diameter InAs nanowires grown by chemical beam epitaxy (CBE).[47] Devices were fabricated on 0.001–0.005 Ω cm As-doped (100) Si wafer (Silicon Valley Microelectronics) with a 100 nm thermal oxide and an additional 10 nm HfO$_2$ layer deposited by atomic layer deposition. This wafer was prepatterned with Ti/Au interconnects and electron beam lithography (EBL) alignment structures before being divided into smaller "chips" on which nanowire transistors were made. Nanowires were deposited by dry transfer using lab wipe. Source, drain and gate electrodes and heaters were defined by EBL using a Raith 150-two system. The EBL resist was a 5% solution of 950 k molecular weight polymethylmethacrylate in anisole (Microchem) deposited by spin coating at 5000 rpm followed by a 5 min hotplate bake at 180 °C. The resist is developed using a 1:3 mixture of methylisobutylketone in 2-propanol. The electrodes and heaters consisted of 25 nm Ni and 75 nm Au deposited by thermal evaporation, immediately after a 120 s $(NH_4)_2S_x$ contact passivation step at 40 °C.[48] Lift-off was performed overnight in N-methyl-2-pyrrolidone at 80 °C. The polymer electrolyte was formed by mixing poly(ethylene oxide) (Aldrich, MW 100 k) and $LiClO_4\cdot3H_2O$ (Aldrich) in a polymer:salt ratio of 10:1 by sonication in 10 mL of methanol. The resulting mixture was left standing at room temperature overnight to precipitate out large particulates, with the supernatant used for deposition. The solution was spin-coated onto the sample at 4000 rpm for 60 s and followed by a 90 °C hot-plate bake for 30 min. EBL was performed at 5 kV accelerating voltage with an electron dose of 100 $\mu C/cm^2$ under high vacuum. High energy electrons cross-link the polymer electrolyte causing exposed regions to become relatively insoluble during development.[14,49] The patterned film was developed in deionized water at room temperature for ≈30 s and dried with $N_2$ gas.[14] Completed devices were packaged in LCC20 ceramic chip carriers (Spectrum) and bonded using an Au ball bonder (Kulicke & Soffa 4500). Atomic Force Microscopy (AFM) studies were performed prior to packaging using a Dimension DI-3000 AFM in tapping mode with Veeco OTESPA7 probes. AFM was performed in cleanroom ambient atmosphere (temperature 20 °C and relative humidity 50–60%).

*Device Operation and Electrical Measurement*: Electrical characterization was performed in an Oxford Instruments Heliox $^3$He cryostat with base temperature $T \approx 275$ mK. The ac source-drain current $I_{sd}$ was measured using a Stanford Research Systems SRS830 lock-in amplifier. The source-drain bias $V_{sd}$ consists of an ac excitation $V^{ac}_{sd} = 100$ μV at a frequency of 13 Hz, with a variable dc component $V^{dc}_{sd}$ added for the measurements in Figure 4b. The dc component is delivered by a Yokogawa 7651 voltage source, with the ac and dc components combined using a passive adder circuit. Yokogawa 7651 voltage sources were used to control both the back-gate bias $V_{BG}$ and the polymer electrolyte gate voltage $V_{PE}$. The dc heater voltage $V_H$ was applied using a Keithley 2400 voltage source enabling continuous measurement of the heater current $I_H$. The applied $V_H$ leads to a temperature gradient along the nanowire via Joule heating in the heater strip as well as an increase in average temperature for the device. The average device temperature was estimated by comparing CB peaks at different fridge temperatures with CB peaks with different applied heating voltages. The resulting dc thermovoltage $V_{th}$ was amplified using a Stanford SR560 voltage preamplifier at a gain of $10^3$ and then read out using a Keithley 2000 multimeter.

## Supporting Information

Supporting Information is available from the Wiley Online Library or from the author.

## Acknowledgements

This work was funded by the Australian Research Council (ARC DP110103802), the Nanometer Structure Consortium at Lund University (nmC@LU), the Swedish Strategic Foundation, the Swedish Energy Agency (project number 38331–1) and Knut and Alice Wallenberg Foundation (KAW). A.P.M. acknowledges an ARC Future Fellowship (FT0990285). D.J.C. acknowledges support from the Australian Nanotechnology Network Overseas Travel Fellowship scheme and the Solander Program. This work was performed in part using the NSW node of the Australian National Fabrication Facility (ANFF) and in part using Lund Nano Lab (LNL).

Received: August 25, 2014
Revised: October 5, 2014
Published online:

---

**Figure 1**

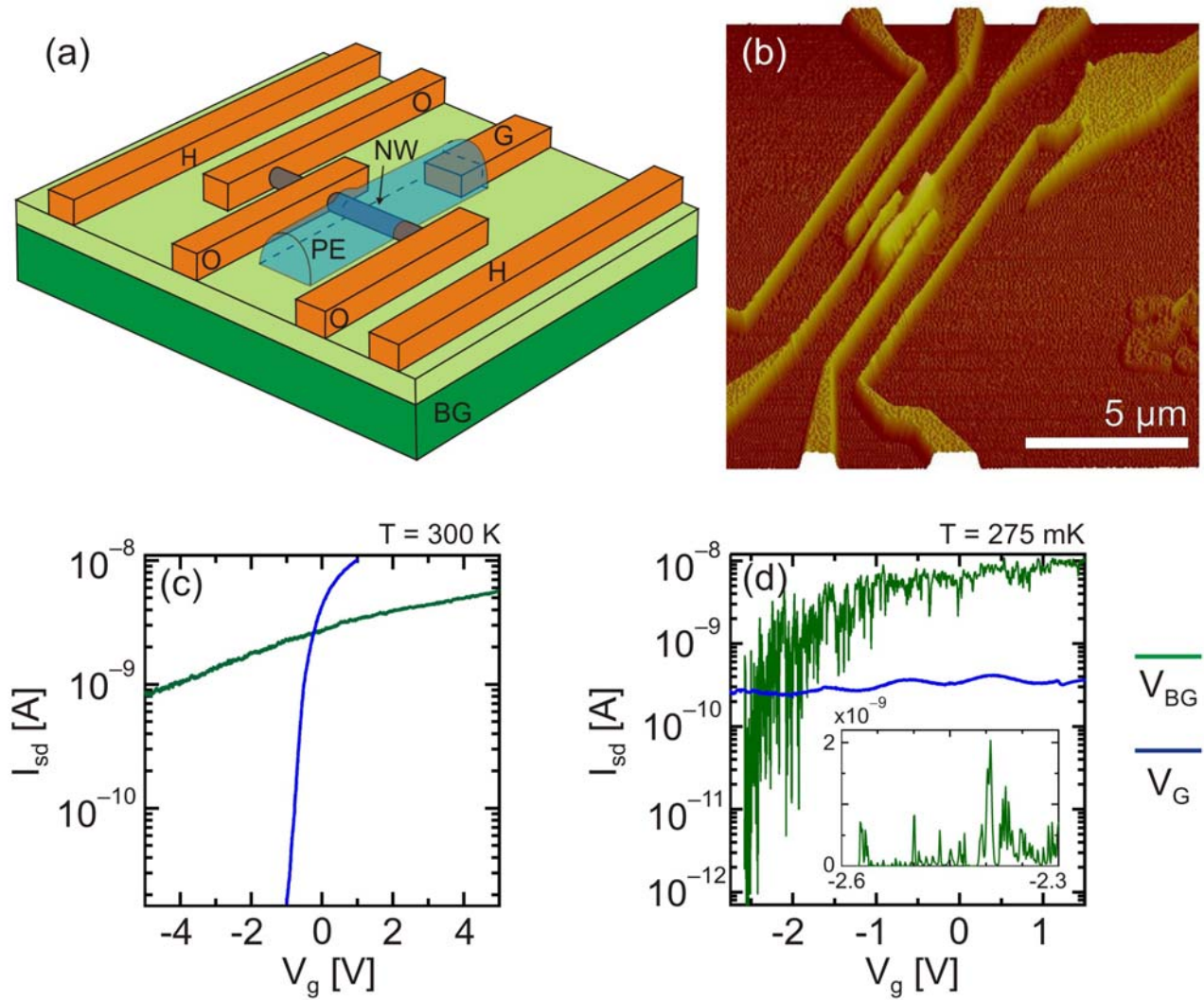

Figure 1: (a) Schematic and (b) atomic force micrograph of the device used in this study, which consists of an InAs nanowire (NW), three NiAu ohmic contacts (O), an insulated n⁺-Si back-gate (BG), a polymer electrolyte dielectric (PE) connected to a gate electrode (G) and two heater strips (H) for applying a thermal gradient along the nanowire. (c, d) Source-drain current $I_{sd}$ vs gate voltage $V_g$ at temperatures (c) $T = 300$ K and (d) $T = 0.275$ K. In each, the green trace is the back-gate voltage $V_g = V_{BG}$ and the blue trace is the PE gate voltage $V_g = V_G$, with the other gate held at ground. In (c) we have $V_{PE} = V_g$ whereas in (d) $V_{PE}$ is fixed at 0 V due to ionic freeze-out. The inset to (d) is a focus on $I_{sd}$ at most negative $V_{BG}$.

**Figure 2**

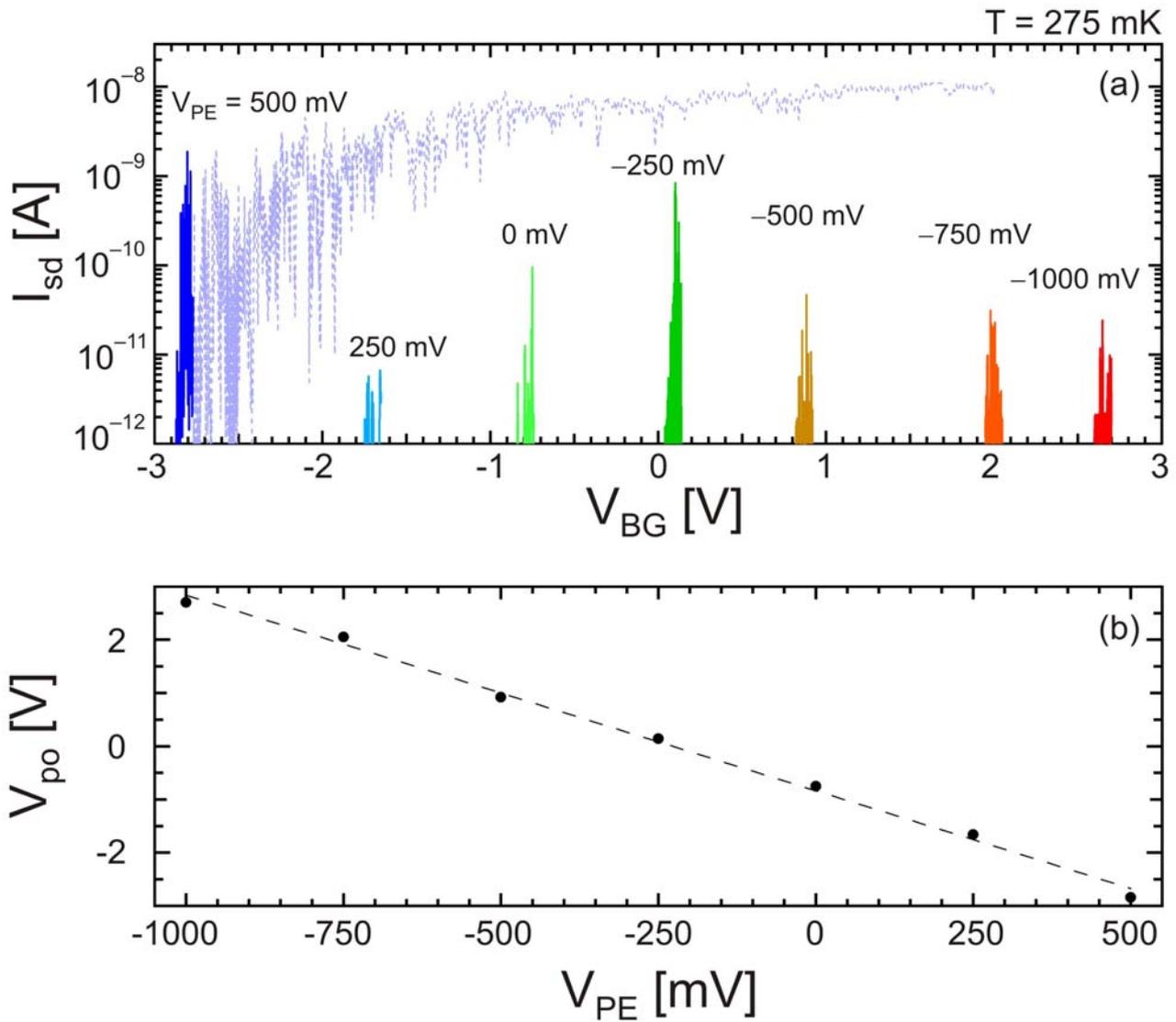

Figure 2: (a) $I_{sd}$ vs $V_{BG}$ for the first 100 mV of $V_{BG} > V_{po}$ for the seven different $V_{PE}$ settings indicated. Additionally, the full $V_{BG}$ range for the gate trace at $V_{PE}$ = 500 mV is shown as the faded blue dashed line in (a). To set $V_{PE}$, the device is warmed to room temperature, the $V_{PE}$ is changed, and the device is cooled back to $T$ = 0.275 K. We define the pinch-off voltage $V_{po}$ as the first $V_{BG}$ where $I_{sd} > 10$ pA if starting from highly negative $V_{BG}$. (b) Plot of $V_{po}$ vs $V_{PE}$ demonstrating the ability to use the PE gate to set the threshold for the underlying n[+]-Si substrate back-gate.

**Fahlvik Svensson *et al.* doi:10.1002/adfm.201402921 – Colour figures.**

**Figure 3**

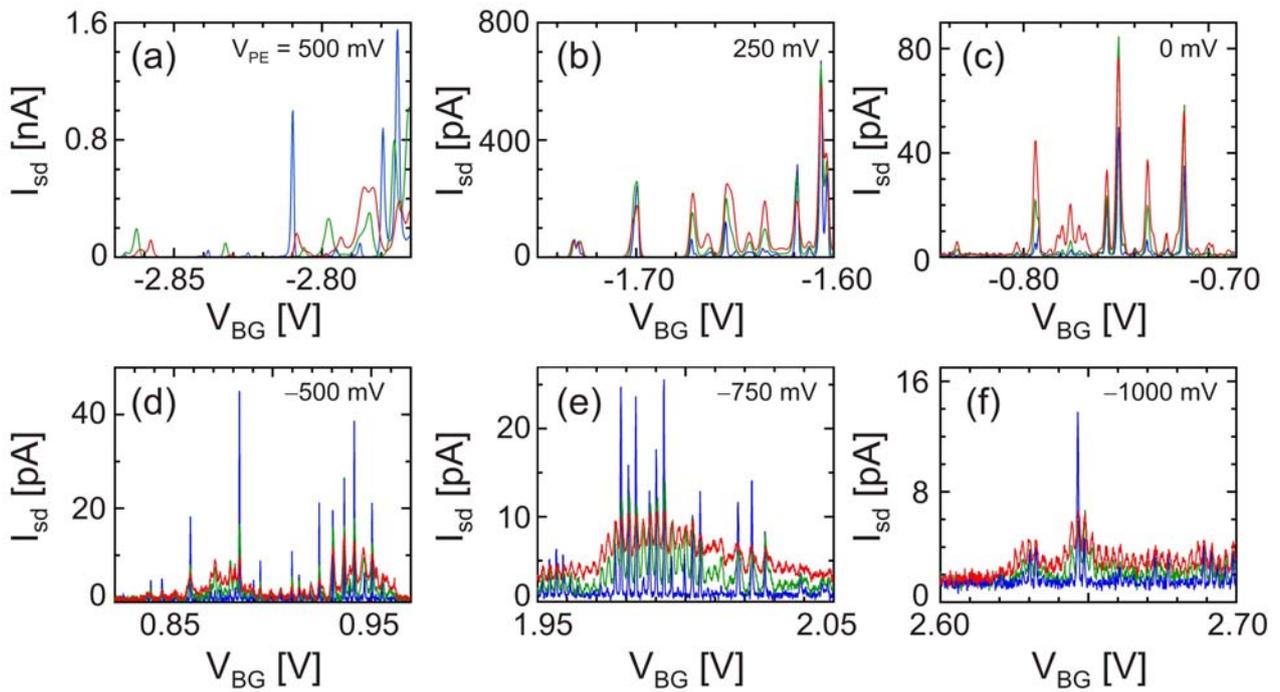

Figure 3: (a-f) $I_{sd}$ vs $V_{BG}$ for $T$ = 275 mK (blue), 1.5 K (green) and 2.5 K (red) for six different $V_{PE}$ settings: (a) $V_{PE}$ = 500 mV, (b) 250 mV, (c) 0 mV, (d) −500 mV, (e) −750 mV, and (f) −1000 mV. The traces show an evolution from highly stochastic Coulomb blockade (CB) at most positive $V_{PE}$ to more regular, periodic CB oscillations at more negative $V_{PE}$.

**Figure 4**

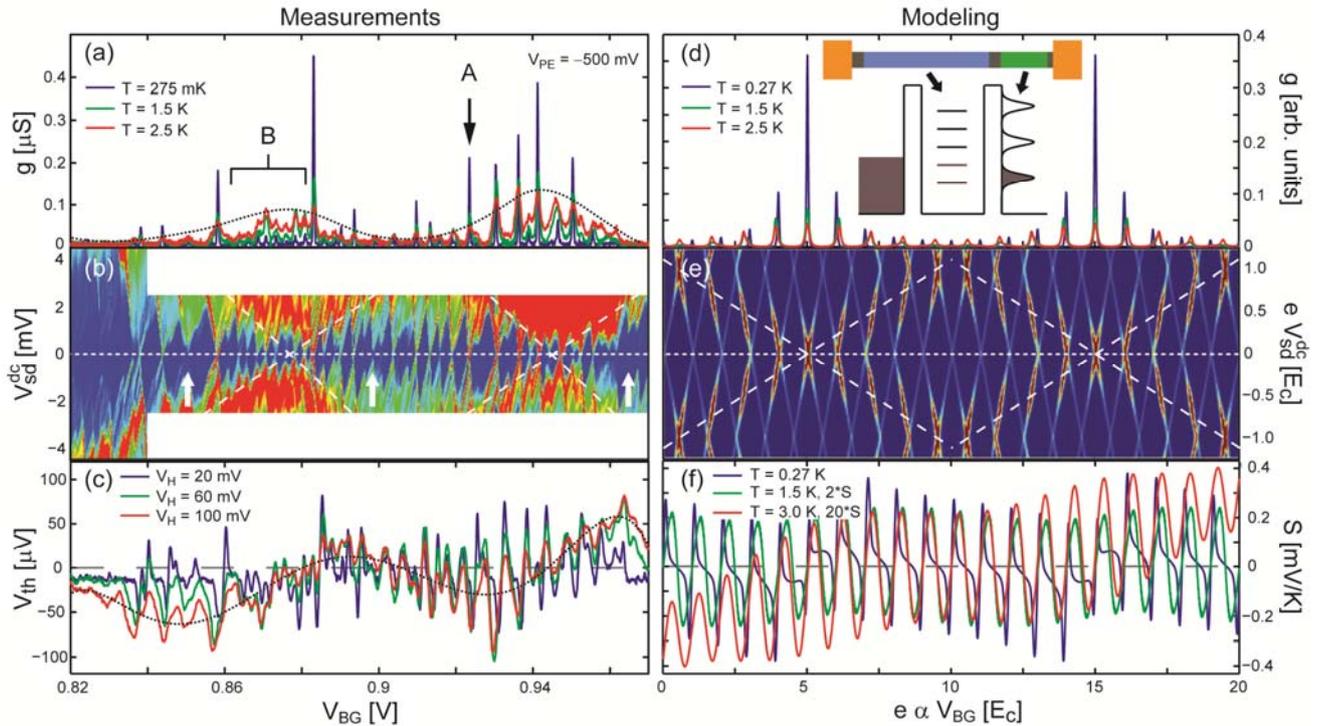

Figure 4: Electric/thermoelectric measurements (a-c) and associated theoretical modeling (d-f) for our NW device with $V_{PE}$ = −500 mV. (a) Differential conductance $g$ vs $V_{BG}$ for three different temperatures; this data matches that in Figure 3(d). A and B point to regions where the peak $g$ decreases/increases with increasing $T$, similar to quantum and classical CB. The dotted line is an approximate envelope for the CB peaks at $T$ = 2.5 K. (b) CB stability diagram showing $g$ (color axis) vs the dc source-drain bias $V^{dc}_{sd}$ ($y$-axis) and $V_{BG}$ ($x$-axis). The color axis runs from $g$ = 0 (dark blue) through to $g \cong 1$ μS (red). The data indicates that the NW splits into two adjacent dots – one large and the other small and strongly coupled to a reservoir. The small blue diamonds correspond to CB in the larger dot, the larger diamonds indicated by the white dashed lines correspond to CB in the smaller dot. The white dotted line indicates $V_{sd}$ = 0 and the white arrows indicate small diamonds which fail to properly close (see text). (c) Open circuit dc thermovoltage $V_{th}$ vs $V_{BG}$ for three heater voltages $V_{H}$. The dotted black line is a $7^{th}$ order polynomial fit to the $V_{H}$ = 100 mV data to highlight the background trend (see text). (d) Calculated $g$ vs effective gate voltage $e$ for the three $T$ values in (a). The inset is a schematic of the theoretical model, which consists of a nanowire quantum dot connected to a conventional reservoir on one side and a reservoir with a modulated density of states on the other. (e) Calculated stability diagram corresponding to (b), the white dashed lines passing through the g peaks correspond to those in (b). (f) Calculated thermopower $S$ vs $V_{BG}$ for three temperatures $T$ (see text).

**ADVANCED
FUNCTIONAL
MATERIALS**

# Supporting Information

Using Polymer Electrolyte Gates to Set-and-Freeze Threshold
Voltage and Local Potential in Nanowire-based Devices and
Thermoelectrics

*Sofia Fahlvik Svensson, Adam M. Burke, Damon J. Carrad,
Martin Leijnse, Heiner Linke, and Adam P. Micolich\**

# WILEY-VCH

## Supporting Information

**Using polymer electrolyte gates to set-and-freeze threshold voltage and local potential in nanowire-based devices and thermoelectrics**

*Sofia Fahlvik Svensson, Adam M. Burke, Damon J. Carrad, Martin Leijnse, Heiner Linke and Adam P. Micolich\**

### Electrical Measurements

The circuits used to obtain electrical conductance and thermovoltage measurements of our device are shown in **Figure S1**. For the conductance measurements a Stanford Research Systems SRS830 lock in amplifier – denoted by the dashed box – was used to source a 100 µV ac voltage $V^{ac}_{sd}$ at a frequency of 13 Hz and measure the resulting ac current $I_{sd}$ at the drain contact via a Femto DLPCA preamplifier. An additional dc voltage $V^{dc}_{sd}$ was applied using a Yokogawa 7651 for the source-drain bias spectroscopy measurements. The ac and dc source-drain bias are added via a simple circuit consisting only of passive components (i.e., a small resistor-divider network and an RC filter to ground to minimise ground loops/noise). We used a Keithley 2400 to apply the gate voltages to electrode G and the substrate, these in turn set $V_G$ (and $V_{PE}$ if $T \gtrsim 200$ K) and $V_{BG}$, respectively.

For the thermovoltage measurements, we used an additional Keithley 2400 to supply a voltage $V_H$ to a resistive heating strip. The associated Joule heating provides the temperature gradient $\Delta T$ that produces an open-circuit thermovoltage $V_{th}$, which is amplified by a Stanford

SR560 voltage preamplifier and detected using a Keithley 2000 multimeter. For the thermovoltage measurements, $V_{PE}$ and $V_{BG}$ were applied as in the conductance measurements.

**Characterization of the polymer gate with temperature**

**Figure S2** shows $I_{sd}$ versus $V_G$ characteristics for the polymer electrolyte gate as a function of temperature $T$. At close to room temperature, the PE gate has a low sub-threshold swing ~ 300 mV/dec. As $T$ is decreased, the polymer gate sub-threshold characteristics worsen rapidly, such that at $T = 220$ K the sub-threshold swing has increased over an order of magnitude. At $T = 220$ K, the gating effect is comparable to that observed for nanowire devices with no gate dielectric,[1,2] and the ions are no longer able to migrate. As a result, for $T < 220$ K, the effective polymer gate voltage $V_{PE}$, as seen by the nanowire, is no longer influenced by any change in the voltage $V_G$ applied to the electrode G for the polymer electrolyte gate. Further discussion of this aspect can be found in the main text, and also in Panzer *et al.*[3]

**Additional data set for the device at negative PE gate voltage**

**Figure S3** shows a data-set matching that in Figure 4 obtained at another, more negative, polymer electrolyte gate voltage $V_{PE} = -750$ mV to demonstrate qualitative repeatability of the reported behavior. This data was obtained from the same device on a separate cool-down. All of the essential behaviors in Figure 4 are reproduced in this case including: i) observation of CB peaks that both increase and decrease in height with increasing $T$ in the $g$ versus $V_{BG}$ data, ii) appearance of small blue diamonds and a larger red/green diamond structure at higher $V^{dc}_{sd}$, and iii) clear correspondence between behavior with $T$ for data in (a), with $V_H$ for data in (c), and structures in (b). Variations are indicative of some morphological changes in the smaller quantum dot, which are expected given that $V_{PE}$ and the resulting ionic environment around the nanowire have changed.

**Further discussion regarding the length estimate for the smaller QD**

The 30 nm length estimate for the smaller QD is at best a lower-bound because the gate coupling measured from the slope of the corresponding diamond edge will be significantly affected by capacitive coupling effects between the QDs[4] – the larger QD occupancy changes by around 20 electrons between two consecutive $V^{dc}_{sd} = 0$ diamond vertices of the smaller QD (see main text Figure 4 (b)). Additionally, our measurement entails some uncertainty as we have a limited length of diamond edge to work with for the small QD. As an upper-bound estimate we can assume the small QD has the same gate coupling as the large QD; in this case we obtain a small QD length of ~ 200 nm. Ultimately, these lengths are reasonably consistent with what we might expect given the device architecture (Figure 1).

**Characterization of how heater voltage affects device temperature**

**Figure S4** shows the data used for an analysis of how the applied heater voltage $V_H$ influences the device temperature $T$. When $V_H$ is applied it results in both a temperature gradient along the nanowire, which produces $V_{th}$, and an overall increase in average temperature $T$ of the device. The device temperature can also be changed using a heater mounted on the cryostat's ${}^3$He pot – this produces a sample-wide temperature increase and no thermal gradient along the nanowire. We can obtain the correspondence between $V_H$ and $T$ by using the Coulomb Blockade (CB) peaks as a 'thermometer'. First we measure how the CB peak heights change with $T$, as controlled by the ${}^3$He pot heater (see Figure S4 (a/c)). Then we measure how the CB peak heights change with $V_H$ at nominally fixed $T$ (see Figure S4 (b/d)). To do this we switch the ${}^3$He pot heater off, let the device settle to base temperature $T = 270$ mK and measure. The ${}^3$He pot temperature does not remain perfectly fixed during this process, it increases from 270 mK at $V_H = 0$ V to 287 mK at $V_H = 0.1$ V. We used the resulting CB peak height graphs (Figure S4 (c/d)) to map, via linear fits, a given $V_H$ to a corresponding $T$. For the three $V_H$ values used in Figure 4, we obtain the following average temperatures: $V_H = 20$ mV gives $T = 0.7$ K, $V_H = 60$ mV gives $T = 1.3$ K, and $V_H = 100$ mV gives $T = 2.0$ K.

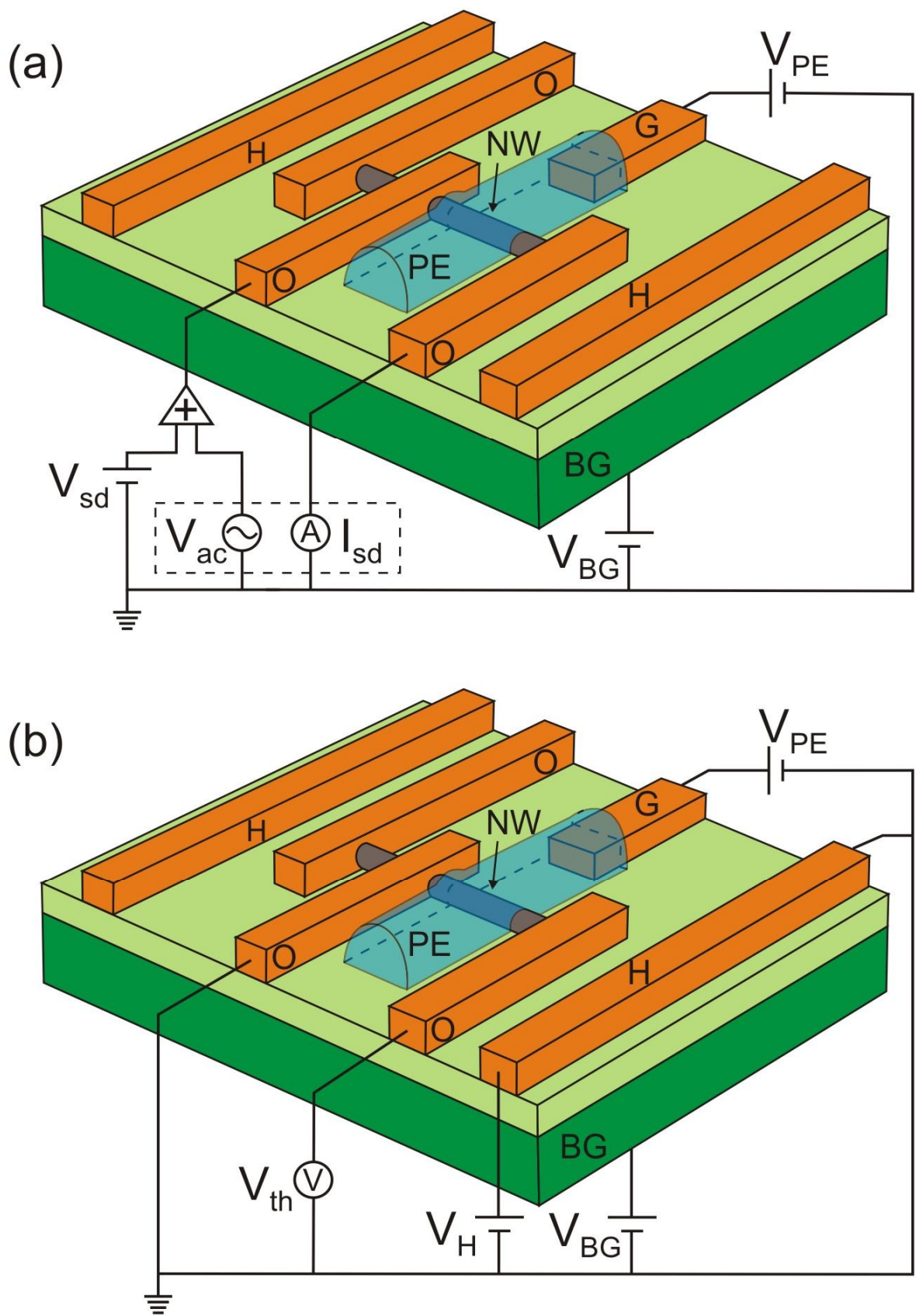

**Figure S1.** The circuits used for (a) electrical conductance and (b) thermovoltage measurements.

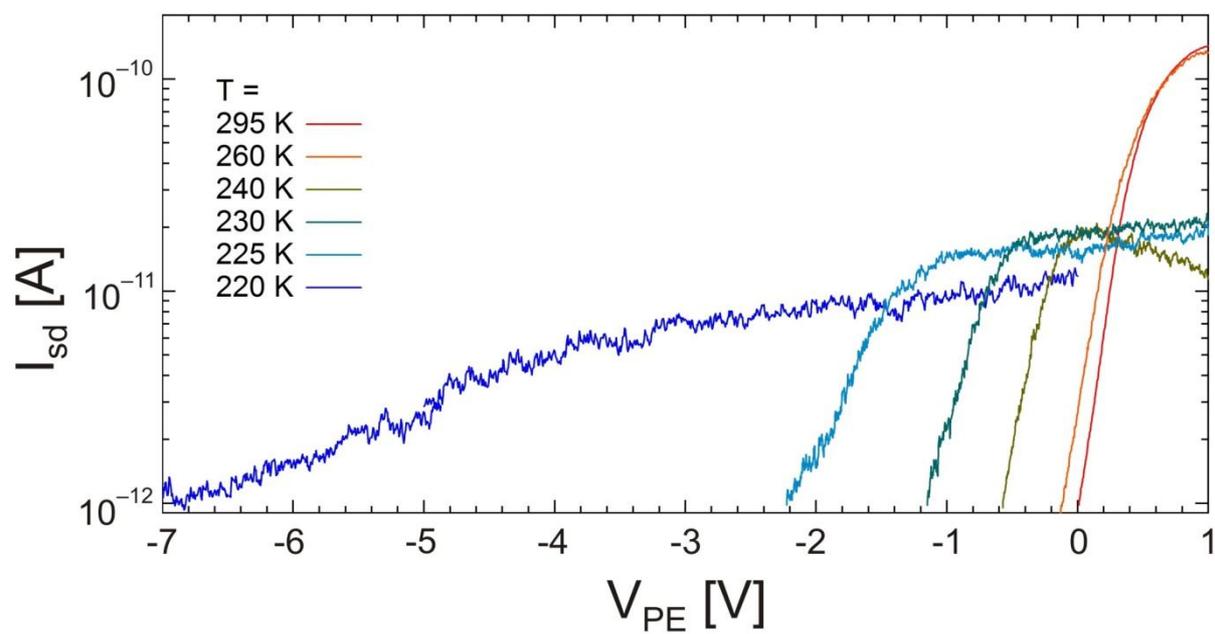

**Figure S2.** Source-drain current $I_{sd}$ vs the voltage $V_G$ applied to the PE gate electrode G for different temperatures $T < 300$ K.

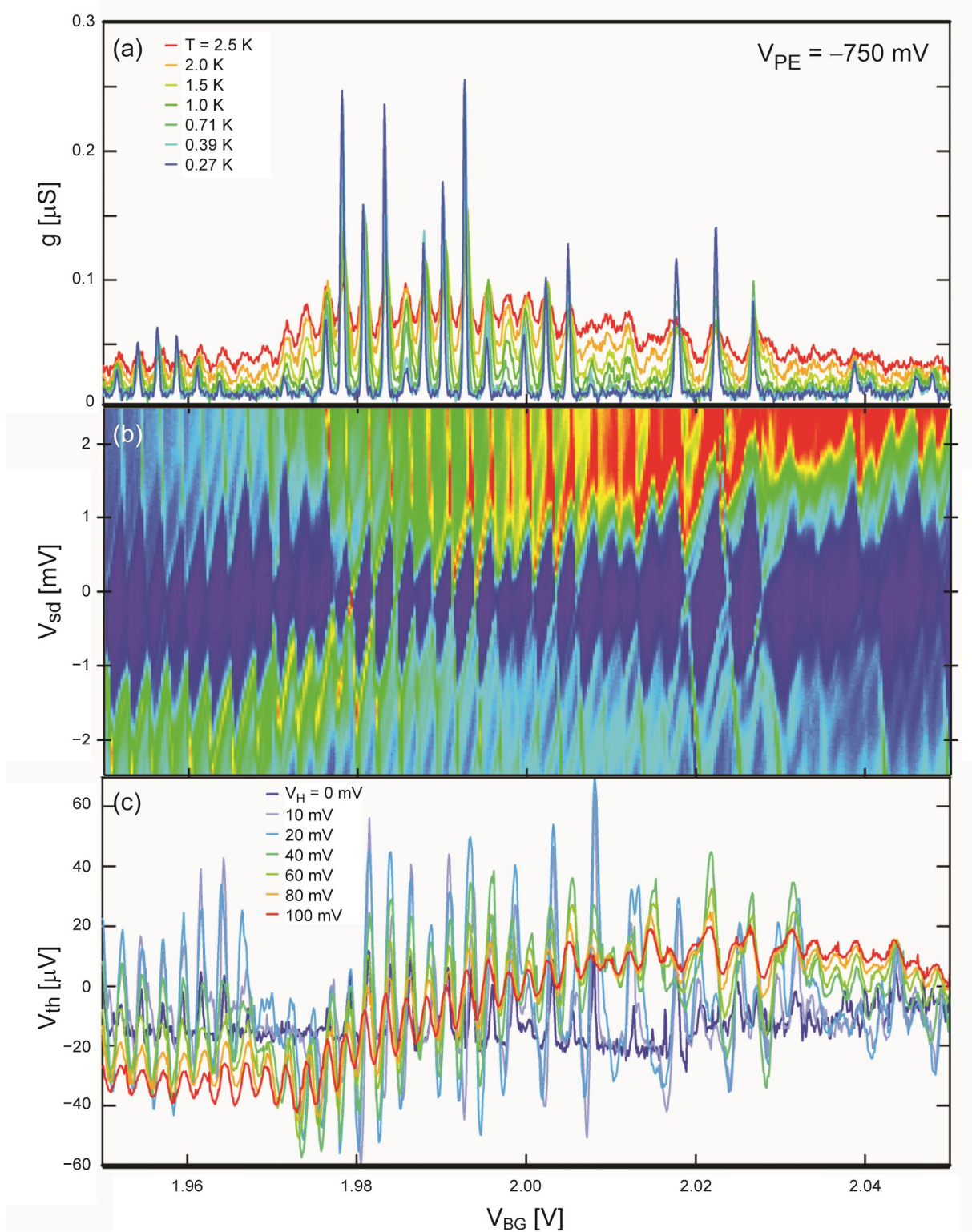

**Figure S3.** (a) Differential conductance *g* vs back-gate voltage $V_{BG}$ for different temperatures *T* at $V_{PE} = -750$ mV. (b) CB stability diagram showing *g* (color axis) vs $V_{sd}$ (*y*-axis) and $V_{BG}$ (*x*-axis) at *T* = 275 mK. The color axis runs from *g* = 0 (dark blue) through to $g \cong 1$ μS (red). (c) Open circuit thermovoltage $V_{th}$ vs $V_{BG}$ for seven heater voltages $V_H$ with the cryostat held at base temperature. There are clear similarities with the data in Figure 4 of the main text, and these are discussed in the supplementary text.

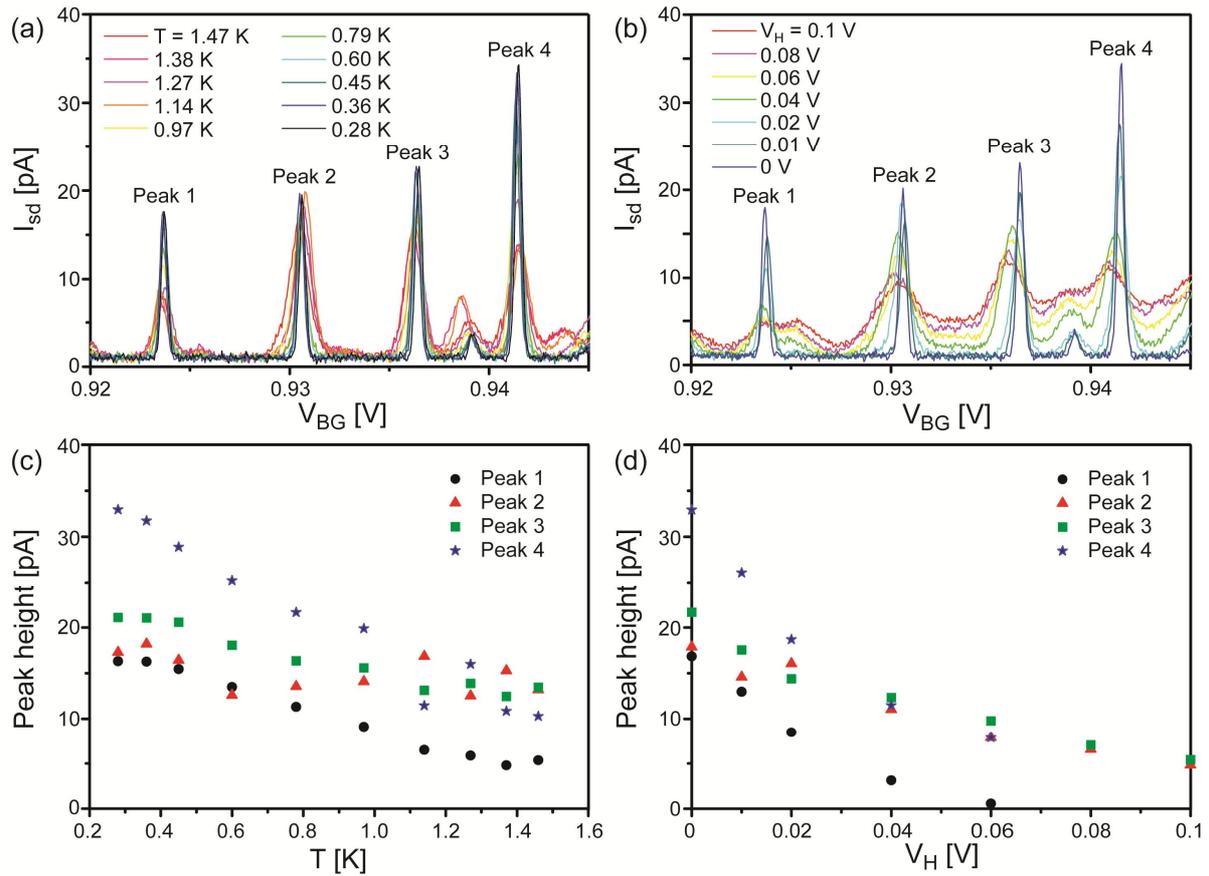

**Figure S4.** (a) Source-drain current $I_{sd}$ vs back-gate voltage $V_{BG}$ at $V_{PE} = -500$ mV for different device temperatures $T$ with $V_H = 0$. (b) $I_{sd}$ vs $V_{BG}$ for the same $V_{BG}$ but for different $V_H$ at $T = 0.28 \pm 0.01$ K. (c) Coulomb Blockade (CB) peak heights extracted from (a) vs $T$. (d) CB peak heights extracted from (b) vs $V_H$.